\documentclass[12pt,preprint]{aastex}

\begin{document}

\title{High Velocity Cloud Edges and Mini-HVCs}

\author{G. Lyle Hoffman}
\affil{Dept. of Physics, Lafayette College, Easton, PA  18042} 
\email{hoffmang@lafayette.edu}

\author{E.E. Salpeter}
\affil{Center for Radiophysics and Space Research, Cornell University, 
Ithaca, NY  14853}
\email{ees12@cornell.edu}

\and

\author{Michael G. Pocceschi}
\affil{Dept. of Physics, Lafayette College, Easton, PA  18042}

\received{}
\accepted{}
\shortauthors{Hoffman, Salpeter \& Pocceschi}
\shorttitle{HVC Edges and Mini-HVC}
\slugcomment{ApJ in press}

\begin{abstract}

Arecibo mapping is reported of the neutral hydrogen distribution along selected 
directions out from the centers of two examples of 
small High Velocity Clouds (HVC).
One HVC (W486) is selected from the class of Compact HVCs (CHVCs)
thought by some researchers to be good candidates for having 
distances characteristic of the Local Group; 
the other (W491) is a bit more extended and possibly nearer.
Both HVCs have a small inner region where the neutral hydrogen column
density $N_{HI}$ decreases slowly and a larger outer region where
$N_{HI}$ declines more rapidly, smoothly and exponentially from 
$\sim 2 \times 10^{19}\;{\rm atoms}\;{\rm cm}^{-2}$ down to 
$< 10^{18}\;{\rm atoms}\;{\rm cm}^{-2}$.
Line widths, and presumably temperature and turbulence, 
do not increase in the outermost regions.
Therefore pressure decreases smoothly, making confinement by dark matter 
gravity more likely than confinement by external pressure.

The more extended HVC, W491, has a superimposed small cloud (which we dub
a ``mini-HVC''), offset 
by $66\;{\rm km}\;{\rm s}^{-1}$ in velocity along the line of sight.
The peak column density of the mini-HVC is about 
$5 \times 10^{18}\;{\rm atoms}\;{\rm cm}^{-2}$.
Preliminary data toward future mapping of two more HVCs reveals two more
mini-HVCs of similarly small size and central column density a bit less 
than $1 \times 10^{19}\;{\rm atoms}\;{\rm cm}^{-2}$, offset by
an even larger velocity, $\sim 98\;{\rm km}\;{\rm s}^{-1}$.
We suggest that these three mini-HVCs are not physically associated
with the HVCs on which they are superimposed, but are either very small
outlyers of the extended Magellanic Stream HVC complex or more distant 
and/or smaller isolated CHVCs.

The importance of the edge column density $N_{1/2}$, the value of $N_{HI}$
at the point where the neutral and ionized column densities are equal,
is discussed.
With $N_{1/2} \sim 2 \times 10^{19}\;{\rm atoms}\;{\rm cm}^{-2}$ for the
two mapped HVCs, the angular scalelength of the total hydrogen is
appreciably larger than the observed \ion{H}{1} scalelength.
Previous distance estimates, related to absolute size and mass of 
the total hydrogen cloud, may have to be scaled down because of the
undetected, more extended ionized hydrogen.

\end{abstract}

\keywords{ISM: atoms --- ISM: clouds; Galaxy: halo --- 
intergalactic medium --- radio lines: ISM}

\section{Introduction}

This paper is mainly concerned with Arecibo\footnote{The Arecibo 
Observatory is part of the National Astronomy and Ionosphere Center, 
which is operated by Cornell University under a management agreement 
with the National Science Foundation.} 
\ion{H}{1} observations 
of two High Velocity Clouds (HVC) and three smaller high velocity features
found serindipitously.
While there is controversy about the location and nature of the various 
types of HVCs \citep{RdBWKGGB99, WHSvWTSBRPK99, PdHS+02, RSW+01}, the 
Compact HVCs (CHVCs) as defined by \citet{BB99} are thought by many 
researchers to be metal-poor 
and associated with the Local Group of galaxies as a whole rather than 
with the immediate vicinity of the Milky Way disk 
\citep{BSTHB99, BB99, BB00, BB01, PdHS+02}.
Even if this hypothesis is correct in general, there is still controversy
as to whether distances to the CHVCs are mostly of order 650 kpc to 1 Mpc
(about the distance to Andromeda) or of order one-third that 
value (still affiliated with the Local Group, but closer than 
its barycenter).
One of the two selected HVCs is a fairly typical CHVC; the other is 
somewhat more extended.
We shall use the term CHVC throughout for clouds such as both of these,
extending the definition of \citet{BB99} to include more extended --- 
and possibly closer --- sources.

The \ion{H}{1} mapping of the two CHVCs reported here gives data relevant to 
three questions:  
(1)  Does the column density decrease roughly exponentially below 
$1 \times 10^{19}$ atoms ${\rm cm}^{-2}$, or is there a sudden change 
in slope somewhere between $2 \times 10^{18}$ and 
$1 \times 10^{19}$ atoms ${\rm cm}^{-2}$?
(2)  Are the cloud edges relatively smooth or severely corrugated 
on $3\arcmin$ scales?
(3)  Does the gas temperature and/or turbulence increase at the 
outermost points, as might be expected if the clouds are surrounded by an 
outside very hot medium?
The present data does not address a fourth question:  
How asymmetric are the clouds?
The three observational questions are related to the following theoretical 
controversies.

A few bright spiral galaxies have been mapped in \ion{H}{1} with good 
sensitivity to large radial distance from the center.
These show an ``outer \ion{H}{1} edge'' in the sense that the 
exponential decrease of the neutral hydrogen column density $N_{HI}$
(as distinct from the {\em total} hydrogen column density $N_{H, tot}$)
shows a sudden steepening beyond some radius \citep{CSS89, vG93}.
The rapid decrease in $N_{HI}$ outside the edge is due to a rapid increase 
in the ionized fraction of hydrogen, in agreement with theoretical models
\citep{CS93, Ma93, DS94}.
We shall denote the theoretical value of the edge column density, defined 
as the neutral gas column density in the region where the neutral and ionized 
fractions are equal, as $N_{1/2}$.
The value of $N_{1/2}$ is close to the observed column density where the
\ion{H}{1} scalelength decreases rapidly, although the exact relation
depends on the assumed morphology of gas and dark matter and 
on the spectrum of the external ionizing flux.
For galaxies $N_{1/2}$ is 
$\sim 2 \times 10^{19}\;{\rm atoms}\;{\rm cm}^{-2}$ and the
central column density is orders of magnitude larger, so the observed
$N_{HI}$ scalelength far inside the edge gives a reliable value for the
$N_{H, tot}$ scalelength everywhere.
Incidentally, observations of Lyman Limit Systems at intermediate 
redshifts give an independent estimate of $N_{1/2}$ \citep{CSB01}, which is
similar (slightly larger than $2 \times 10^{19}$~atoms~${\rm cm}^{-2}$,
but the UV flux at that time was also larger).

Qualitatively similar arguments must apply to the hydrogen in CHVCs
(earlier claims of some ``sharp'' edges on $10$-$20\arcmin$ 
scales are reviewed by \citet{WvW97}).
However, the value of $N_{1/2}$ (and the observed edge column density,
close to that value) depends on gas density and on the ionizing flux,
and could therefore be different for CHVCs.
If $N_{1/2}$ were appreciably smaller for a CHVC than for a galaxy
(as in question 1 above) one could again get the $N_{H, tot}$ scalelength
from the observed $N_{HI}$ scalelength.
However, we shall see that 
$N_{1/2} \sim 2 \times 10^{19}\;{\rm atoms}\;{\rm cm}^{-2}$
is also typical for CHVCs, but that the central column density is
larger by only a factor of a few and the small intervening region is
further complicated by the transition from cold  to warm neutral
hydrogen \citep{BB00}.
As a consequence, the observed \ion{H}{1} scalelengths and masses 
may or may not greatly underestimate the size and mass of total hydrogen.
Numerical modelling will be needed to settle this issue, but we give
some conjectures in Sect. 5.

\citet{MLS95} and \citet{LMPU02} (We shall refer to these papers
jointly as the ``Murphy-Lockman Survey,'' abbreviated MLS.) 
report that 37\% of sightlines to quasars (randomly selected 
with respect to known HVCs) show \ion{H}{1} emission at velocities 
typical of HVCs, with column densities 
$7 \times 10^{17}$~atoms~${\rm cm}^{-2}$ and above when averaged 
within the $21\arcmin$ beam of the Green Bank 43m telescope.
Such column densities are far below the suggested $N_{1/2}$ 
$\sim 2 \times 10^{19}$ atoms ${\rm cm}^{-2}$, and most of the
hydrogen must be ionized if the column density is 
uniform over the $21\arcmin$ beam.
While most of the MLS sources are evidently extended outskirts of 
major HVC complexes,
a few of the more isolated sources might be highly clumpy, 
consisting of much smaller 
``mini-HVCs'' which have central column densities closer to $N_{1/2}$ 
than to $7 \times 10^{17}$~atoms~${\rm cm}^{-2}$ but which cover 
much less than the full $21\arcmin$ beam.
We hope to map (with Arecibo resolution) some of the more isolated MLS 
sightlines in the future, but serendipitously our present observations 
have found three ``mini-HVCs'' with angular sizes larger than the 
Arecibo beam but smaller than $21\arcmin$.

Our Arecibo mapping procedures are detailed in Sect. 2.
Results are presented in Sect. 3 for W486 and W491 and in Sect. 4 for
the three mini-HVCs.
The CHVC distance controversy is discussed in Sect. 5,
along with the relation of the mini-HVCs to that controversy, 
likely values of $N_{1/2}$, and the question of pressure vs.\ gravity 
confinement.
Sect. 6 restates our conclusions succinctly.

\section{Arecibo Observations}

We selected two HVCs from the compilation of \citet{WvW91} for study:  
one object, entry number 486 (hereafter referred to as W486) in that 
list, is considered a typical Compact HVC (CHVC) by \citet{BB99}; a map 
at the resolution of the Dwingeloo 25 m radio-telescope was given there.
The second HVC, W491, is somewhat more extended and possibly nearer, 
but more likely to have properties similar to the CHVC than to the large
complexes.
Throughout this paper we will broaden the definition of the term 
CHVC beyond the strict definition
of \citet{BB99} to include somewhat larger and somewhat less 
isolated HVCs like W491 --- all HVCs that are not more than $3\arcdeg$ 
in diameter as listed in \citet{WvW91} and not clearly associated with
any of the major complexes.
A map of W491 using the NRAO 43m telescope was given by \citet{Wr79}.
W486 is part of Wakker \& van Woerden's Anti-Center Very High Velocity 
(ACVHV) complex in Population AC (clouds near the Galactic anticenter 
with $v_{LSR} < -150\;{\rm km}\:{\rm s}^{-1}$) while W491 is a member 
of Population EN (extreme negative-velocity clouds that are not 
associated with the Magellanic Stream nor the Galactic Center nor 
Anti-Center populations).
Both have $v_{LSR} < -250\;{\rm km}\:{\rm s}^{-1}$, well removed 
in velocity from emission from the Galactic disk.
Both are at southern Galactic latitudes and were observed during 
nighttime at Arecibo in order to have the quietest possible spectral 
baselines.

Observations were conducted in July/August 1999 and August 2000.
We used the Gregorian feed system with the ``L narrow'' receiver 
in total power (position-switched) mode, with 3.05 kHz 
(about $0.63\;{\rm km}\:{\rm s}^{-1}$) channel spacing.
The reference beams trailed the on-source beam positions by $6^m$
in right ascension ($\sim 85\arcmin$ at the declinations of these
sources), in all cases well outside the contours of the clouds in
the available lower resolution maps.
Calibration was accomplished by observing several continuum sources 
from the VLA calibrator list, chosen to have small size compared to 
the $3\farcm2$ beam.
In addition, we reobserved a few spiral galaxies for which we had 
high signal-to-noise pre-upgrade \ion{H}{1} measurements and which 
were known to be $\ll 3\farcm2$ in extent.
After calibration, each spectrum was baselined and smoothed with a 
Savitsky-Golay filter. 

Short-integration mapping a few beamwidths around the nominal HVC 
centers served to locate the highest column density cores of the 
clouds more precisely.
Armed with that information, we proceeded to map along a straight 
line (i.e., a ``spoke,'' analogous to a spoke of a wheel) 
from the center toward the nearest edge of 
each cloud (as indicated in the coarser maps 
cited above) with beam positions separated by a beamwidth.
Integration times ranged from 5 min ON at inner positions to 50 min ON at the 
outermost positions, approximately $30\arcmin\:$ from the center.
To set limits on possible sidelobe contributions, we first mapped 
several bright continuum sources of 
negligible extent at declinations comparable to those of our two HVC, 
then observed beam positions with perpendicular separation 
$3\farcm2 \sqrt 3 = 5\farcm54$ from the positions along the spoke,
as indicated on Fig. 3 for W486.
(Fig. 5 shows the points to the west of the spoke for W491; 
corresponding points to the east are not shown.)
This placed the off-spoke points near the peak of the first sidelobe 
ring for the next outermost point on the chosen spoke.
In the follow-up mapping, conducted in August 2000, we added more points 
off-spoke in both directions for W486 (shown in Fig.\ 3 below), 
to investigate how corrugated 
the edge might be at the $2 \times 10^{18}$ atoms ${\rm cm}^{-2}$ 
level, and we mapped the full extent of a cloudlet (``mini-HVC'' in
the language of the Introduction) discovered 
superimposed on the western off-spoke points for W491 (see Fig.\ 5 below).

The sidelobe was found to peak at a level of $-15$ dB on average, 
in a slightly elliptical ring about $10 \times 11\arcmin\;$ in 
diameter, with the minimum between the main beam and the sidelobe 
being similarly elliptical, about $7 \times 8\arcmin\;$ in diameter
(see also \citet{HPN+01}).
The strength varies considerably with azimuth and zenith angle, 
however; in some orientations the maximum sidelobe response is as 
high as $-12$ dB, in others the response is below $-18$ dB.
On average (appropriate when repeated scans on each point occur 
in different orientations of the Gregorian subreflector), each 
octant of the sidelobe ring contributes about 4.7\% of the response 
of the main beam, each sextant about 6.2\%.
For a complete map, these figures would allow us to approximately 
subtract sidelobe contributions iteratively.
Since we do not have complete maps of these clouds, we estimated sidelobe 
contributions by assuming approximate symmetry about the centers of the clouds.
In this way, we determined that the first sidelobe ring contributes 
about 20\% to each spectrum (i.e., about half the response it would 
give for a completely uniform source much more extended than the 
sidelobe ring) for these observations.
Since reduction of each flux by a fixed percentage does not affect the
derived scalelength, and to avoid compounding uncertainties, we present 
the spectra without any correction for sidelobes.

``Far'' sidelobes (i.e., all those beyond the first sidelobe ring discussed 
above, including what is sometimes called ``stray radiation''), due to 
scattering off the triangular platform and the 
support cables as well as diffraction through the partially blocked 
aperture, are known to exist \citep{PH01}.
In at least one pathological orientation of the Gregorian subreflector, 
there is a spot in the ``far'' sidelobes which rises almost to the level 
of the first sidelobe ring.
However, in most orientations the ``far'' sidelobes are down by at least 
another 3 or 4 dB from the first sidelobe ring.
The average integrated response of the ``far'' sidelobes is unknown at 
this point.
For observations on the outskirts of a very strong source like DDO~154, 
``far'' sidelobes can contribute noticeably \citep{HvGSC01} in integrations 
as long as the ones employed here.
The core of W491 is weaker by a factor of 5 than DDO~154, that of W486 
by a factor of 10.
That should reduce the ``far'' sidelobe contributions to the noise level 
in these observations, except in pathological orientations.

\section{Results for the Targetted HVCs}

\subsection{W486 = CHVC 39 = HVC 158-39-285}

Representative spectra for the CHVC W486 (with no correction for 
sidelobes) are shown in Figure 1.
The figure includes a spectrum (labelled W486C --- C for ``center'') 
from the core of the cloud, 
integrated 5 min ON, and one (labelled W486R7 --- R7 for ``seventh beamwidth
out along the radius'') from a point 
$7 \times 3\farcm2 = 22\farcm4$ out from the core along the mapped spoke.
The latter was integrated 50 min ON and has rms noise 0.7 mJy after 
smoothing, typical of the outermost spectra on each cloud.
The displacement in velocity by $17\;{\rm km}\;{\rm s}^{-1}$ is 
characteristic of the trend along the spoke:  profile center velocities 
(defined to be the midpoint between the points at 50\% of peak on either 
side of the feature) rise monotonically from $-285\;{\rm km}\;{\rm s}^{-1}$ 
(heliocentric; $\upsilon_{LSR} = -292$, $\upsilon_{GSR} = -228$, 
$\upsilon_{LGSR} = -150\;{\rm km}\;{\rm s}^{-1}$) at the core to 
$\upsilon_{\sun} = -268\;{\rm km}\;{\rm s}^{-1}$ at the 
outermost points.
Whether this represents systematic rotation or a shear (tidal or 
otherwise) remains to be determined, but \citet{BB00} report similar 
velocity gradients for the cores of the six CHVC they have mapped 
at Westerbork.
Velocity corrections to GSR and LGSR follow the conventions of \citet{BB99}.
Profile widths (FWHM) remain essentially constant along 
the spoke, ranging between 23 and $27\;{\rm km}\;{\rm s}^{-1}$ for 
the core through R6.
The outer, noiser points appear slightly broader (The spectrum from a
point 7 beamwidths out from center measures 
$34\;{\rm km}\;{\rm s}^{-1}$ FWHM) but that is 
likely to be an artifact of the noise.

The neutral hydrogen column density along the mapped spoke is shown
in Fig.\ 2.
It is fairly constant near $3 \times 10^{19}\;{\rm atoms}\;{\rm cm}^{-2}$
for about $13\arcmin$ out from the center, then tails off smoothly 
and exponentially from $\sim 2 \times 10^{19}$ down to 
$\sim 1 \times 10^{18}\;{\rm atoms}\;{\rm cm}^{-2}$ at a radius of 
$\sim 30\arcmin$.
Sidelobe corrections, if we could apply them accurately, might shift
all points downward by 0.10 in $\log N_{HI}$ with no significant change
in slope.
The scalelength is $\sim 5\farcm5$.
That is a bit smaller than might be inferred from the constant 
declination cut presented by \citet{BBC01a}, 
consistent with our choice of mapping along the spoke with the 
greatest gradient in column density as suggested by the map 
of \citet{BB99} and confirmed by the more detailed map of \citet{BBC01a}.
The steady shift in centroid velocity of our profiles is consistent 
with the velocity field for the higher column density parts of the 
cloud in \citet{BBC01a}.
There are no signs of the outer edge being broken up into cloudlets 
on the scale of the Arecibo beam ($3\farcm2$).

The fluxes, integrated over velocity, for each observed beam position 
are shown in Fig.\ 3.
Each observed point has been convolved with a $3\farcm2$ beam, 
and each beam position is marked by a small cross.
The greyscale is logarithmic.
Some asymmetry perpendicular to the chosen spoke is evident, and the edge 
is not perfectly smooth, but the beam-to-beam variations perpendicular to 
the spoke do not exceed a factor of a few.

\subsection{W491 = HVC 102-40-409}

Representative spectra for W491 (no correction for sidelobes) are shown in 
Fig.\ 4.
The centroid velocity is $\upsilon_{\sun} = -415$, $\upsilon_{LSR} = -412$,
$\upsilon_{GSR} = -249$, $\upsilon_{LGSR} = -186\;{\rm km}\;{\rm s}^{-1}$.
In this case no systematic velocity gradient is evident along the chosen spoke 
although the profile width increases slightly, from 
$15\;{\rm km}\;{\rm s}^{-1}$ at the center of the cloud to 
$\sim 30\;{\rm km}\;{\rm s}^{-1}$ at a point $16\arcmin$ out, and then 
remains approximately constant at $\sim 30\;{\rm km}\;{\rm s}^{-1}$ from 
that point on out.
The narrower profiles at the core of the cloud are consistent with there
being a cold core similar to those observed with synthesis arrays 
for other CHVCs \citep{BB00}.

The fluxes, integrated over velocity separately for the main cloud and 
the mini-HVC discussed below, are shown in Fig.\ 5.
Convolution and greyscale are the same as for W486 (Fig.\ 3).
For the main cloud we were not able (due to time constraints) to map to as 
low a column density over as wide a range of positions as for W486, but as 
best we can tell there are no significant corrugations of the portion of 
the edge that we have mapped.

The column density vs.\ radius for the main cloud of W491 is shown 
in the righthand panel of Fig.\ 2.
The central region has nearly constant column density, with
a slow decline from $\sim 5 \times 10^{19}$ to 
$\sim 2 \times 10^{19}\;{\rm atoms}\;{\rm cm}^{-2}$.
After that the decline is more rapid; it is not quite so smooth as 
for W486, but it is 
still best characterized as exponential from $\sim 2 \times 10^{19}$
down to the lowest column density observed, just less than 
$1 \times 10^{18}\;{\rm atoms}\;{\rm cm}^{-2}$, with a slightly 
larger scalelength $\sim 7\arcmin$.
In detail, on the chosen spoke the decline is more nearly exponential, with a 
scalelength of $\sim 5\arcmin$, to 
$\sim 2 \times 10^{18}\;{\rm atoms}\;{\rm cm}^{-2}$, then there is a 
$6\arcmin\:$ wide plateau before the decline resumes with 
approximately the same scalelength as before.
A better sense of the decline can be gotten from Fig.\ 6, in which the 
column density for the main cloud alone (i.e., excluding the velocity 
range in which the mini-HVC appears) is plotted as a function of the 
distance from our fiducial cloud center as if the cloud were spherical.
Here the point to point variations wash out the ``plateau,'' leaving 
on average a simple exponential decline with a scalelength of about $6\arcmin$.
There are no indications of corrugation of the edge of W491 either.
As for W486, sidelobe corrections would shift all data points downward by
0.1 in $\log N_{HI}$ without significantly changing the derived scalelength.

\section{Low Column Density Mini-HVCs}

\subsection{A Mini-HVC Superimposed on W491}

Points to the west of the spoke mapped in W491, observed to check sidelobe 
contamination, exhibited emission displaced $66\;{\rm km}\;{\rm s}^{-1}$ 
higher in velocity.
The emission shows up as a distinct feature (see Fig.\ 4, righthand panel)
in every spectrum in which it appears since both the main cloud 
and the mini-HVC have quite narrow profiles 
($\sim 23$ and $36\;{\rm km}\;{\rm s}^{-1}$ at 
50\% of peak, respectively, at positions where the mini-HVC appears).
We can trace the signal contiguously over a roughly elliptical shape, 
$\sim 20 \times 10\arcmin\;$ in extent, aligned very nearly NS as shown 
in the righthand panel of Fig.\ 5.
There is no trace of signal at this position and velocity in the 
less-resolved and less sensitive Leiden/Dwingeloo Survey \citep{HB97}.
Nor does it appear in the map published by \citet{Wr79}, based 
on NRAO 43m data.
We shall call this small cloud, displaced in velocity from the chosen
CHVC, a ``mini-HVC'' no matter whether it is a part of some other extended
HVC complex or a distinct smaller or more distant CHVC.

Column density as a function of radius is shown for the mini-HVC in Fig.\ 7.
The center is taken to be the point for which the spectrum is shown 
in Fig.\ 3 (right-hand panel), which is close to the point of 
highest column density in the cloudlet.
Spokes in different directions, spaced by $60\arcdeg$, are shown 
with different symbols.
Upper limits are shown in the figure by downward arrows.
Our spatial resolution is poor, but as best we can tell 
the column density falls exponentially in all directions, from a peak
$\sim 5 \times 10^{18}$ to 
$\sim 5 \times 10^{17}\;{\rm atoms}\;{\rm cm}^{-2}$ in 8 or $10\arcmin$.
The \ion{H}{1} scalelength along the major axis is thus of order 
$4\arcmin$, slightly smaller than for the CHVCs W491 or W486 themselves.
Both the peak column density and the extent of the mini-HVC are much
less than those for the CHVCs, so sidelobe corrections should not
exceed the noise in the spectra for the mini-HVC.

\subsection{Two More Mini-HVCs}

In preliminary data for future mapping of two more small HVCs, we have 
uncovered 
two additional mini-HVCs, both superimposed on W479 = HVC 132-37-334.
The peak column densities are 
$\sim 1 \times 10^{19}\;{\rm atoms}\;{\rm cm}^{-2}$ for each, a bit higher 
than for the mini-HVC superimposed on W491.
The angular displacements from the center of W479 are $17\arcmin$ and 
$43\arcmin$; the angular sizes are comparable to that of the mini-HVC 
superimposed on W491 although we will need mapping with longer integration
at each point to know the full extents.
Each has heliocentric velocity $-245\;{\rm km}\;{\rm s}^{-1}$
($\upsilon_{LSR} = -247$, $\upsilon_{GSR} = -118$, 
$\upsilon_{LGSR} = -40\;{\rm km}\;{\rm s}^{-1}$), 
$98\;{\rm km}\;{\rm s}^{-1}$ above the velocity of W479 itself.
The profile widths (at 50\% of peak) are about 
$30\;{\rm km}\;{\rm s}^{-1}$, comparable to that of the W491 mini-HVC.
The preliminary mapping of W413, the second of those two additional HVCs,
did not reveal any mini-HVCs with peak column density 
$\gtrsim 1 \times 10^{19}\;{\rm atoms}\;{\rm cm}^{-2}$.

Fig.\ 8 plots heliocentric velocity as a function of the angle $\theta$ 
between the line-of-sight and the direction to the Local Group center, for our
four CHVCs and the three mini-HVCs.
The four CHVCs were chosen for their particularly large negative 
velocities (and, indirectly, small $\theta$).
The three mini-HVCs all have appreciably less negative velocities than
the CHVCs, and we think it is likely that they are chance superpositions
rather than physical associations.
The radial components of their velocities are close to that for the end of
the Magellanic Stream (which \citet{LMPU02} and \citet{SDKB02} 
argue is $\sim 10 \arcdeg$ more extended than previously thought), 
but also close to that of the Local Group barycenter,
with deviations of only $-40$, $-40$, and $-120\;{\rm km}\;{\rm s}^{-1}$,
respectively.

The area on the sky of each mini-HVC is about one-tenth that of a 
typical CHVC ($\sim 20 \arcmin \times 10 \arcmin$ vs.\ 
$\sim 60 \arcmin \times 30 \arcmin$ or more).
The average of the peak column density is about one-fifth that of an
average CHVC, $\sim 8 \times 10^{18}$ vs.\ 
$\sim 4 \times 10^{19}\;{\rm atoms}\;{\rm cm}^{-2}$.
The three mini-HVCs in total cover about 10\% of the area we have mapped 
so far in studying the four HVCs.
As mentioned, these three cloudlets could be either distant 
and/or smaller CHVCs
or some small structures associated with extensions of the Magellanic Stream.
The latter possibility is perhaps more likely, partly because of the 
low central density and partly because of the greater extension of the 
Magellanic Stream mentioned above.

\section{Discussion}

\subsection{Edge Column Density and Ionized Hydrogen}

We have defined the edge column density, $N_{1/2}$, as the column density 
of neutral atomic hydrogen where the neutral and ionized column densities
are equal.
As mentioned in the Introduction, 
$N_{1/2}$ is of order $2 \times 10^{19}\;{\rm atoms}\;{\rm cm}^{-2}$
for nearby spiral galaxies, as observed by the shortening of the
$N_{HI}$ scalelength when $N_{HI} < N_{1/2}$ and confirmed by theoretical
modelling \citep{CS93, Ma93, DS94}.
Theoretical models have also been carried out for gas clouds representing
Lyman Limit Systems \citep{CSB01, CB02}.
The middle curve of Fig.\ 3 in \citet{CSB01} shows a likely 
$N_{HI} - N_{H, {\rm tot}}$ relation.
Their analysis of the distribution function for $N_{HI}$ 
at intermediate redshifts gave a value of $N_{1/2}$ only slightly larger
than for nearby galaxies, which may reflect the increase of the
extragalactic UV flux with redshift.
As shown in Fig. 2 for the two CHVC main clouds which we mapped in detail, 
W486 and W491, there is a change of slope near 
$N_{HI} \sim 2 \times 10^{19}\;{\rm atoms}\;{\rm cm}^{-2}$
in the \ion{H}{1} column density as a function of radius, and
$N_{HI}$ drops smoothly and approximately exponentially from
$\sim 1 \times 10^{19}$ to $2 \times 10^{18}\;{\rm atoms}\;{\rm cm}^{-2}$.
Discounting the unlikely possibility of 
$N_{1/2} < 2 \times 10^{18}\;{\rm atoms}\;{\rm cm}^{-2}$,
a value of $N_{1/2}$ near $2 \times 10^{19}\;{\rm atoms}\;{\rm cm}^{-2}$
is thus likely for CHVCs, similar to that for galaxies and 
Lyman Limit Systems.

With $N_{1/2} \sim 2 \times 10^{19}\;{\rm atoms}\;{\rm cm}^{-2}$
the hydrogen at the center of W486 and W491 is mostly neutral, 
but further out the undetected contribution of 
ionized hydrogen is uncertain and likely to be large.
In the absence of explicit numerical modelling, we have estimated
$N_{H, tot}$ as follows:
We adopt the $N_{HI} - N_{H, {\rm tot}}$ relation for almost
spherical clouds from \citet{CSB01}, using the middle curve of their 
Fig. 3 which corresponds to 
$N_{1/2} \sim 2 \times 10^{19}\;{\rm atoms}\;{\rm cm}^{-2}$.
We can then construct a curve for $N_{H, tot}$ vs. radius from our
Fig. 2 or 6 out to radius $30\arcmin$.
With these theoretical assumptions, we estimate that the scalelength for
total hydrogen is 3 or 4 times larger than the observed scalelength for
\ion{H}{1} ($\sim 20\arcmin$ for W491).
Allowing for the mostly neutral core (radius $< 10 \arcmin$), we estimate 
a larger total mass $M_{H, {\rm tot}} \sim 5 M_{HI}$.
For comparisons with structure formation calculations, the ratio of baryon
mass to dark mass is important.
This is now $\sim 5$ times larger than estimates in the literature which use
only $M_{HI}$.
For comparison with Lyman Limit Systems, the radius where $N_{HI}$ drops to 
$\sim 1 \times 10^{17}\;{\rm atoms}\;{\rm cm}^{-2}$ is important
\citep{CCR00}.
Using the same $N_{HI} - N_{H, {\rm tot}}$ relation, we estimate this 
to be $\sim 40 \arcmin$
compared with the observed radius $\sim 30 \arcmin$ in Fig.\ 2.

The increases in scalelength and mass to account for the unobserved 
ionized hydrogen, as estimated above, are uncertain and may not be
quite as large for typical CHVCs which tend to have larger central 
$N_{HI}$ than W486 and W491, but the increases are still likely 
to be appreciable.
The earlier papers on the CHVC distance scale 
\citep{BSTHB99, BB99, BB00, BBC01a, BBC01b} showed that typical distances 
$D \sim 600\;{\rm kpc}$ to 1 Mpc led to physical scalelengths 
and baryon fractions consistent with those for dwarf galaxies
if ionized hydrogen were neglected.
The scalelength and baryon fraction both scale as $D$, so the
increase from \ion{H}{1} to ${\rm H}_{tot}$ will give consistency for
somewhat smaller distances $D$.
The scarcity of starless gas clouds in other galaxy groups
\citep{Zw01, ZB00, dBZDBF02} has been used as an argument against
$D \gtrsim 600$ kpc.
The shortening of the distance scale, along with some recalibration,
tends to remove the discrepancy \citep{BB01,PdHS+02}.
Similarly, arguments about distant CHVCs contributing too many
Lyman Limit Systems \citep{CCR00} are ameliorated by a shorter
distance scale.

\subsection{Confinement, Clumpiness, and Other Uncertainties}

There is still some controversy as to whether HVCs, CHVCs in particular,
are confined by dark matter gravity or by pressure from an external
very hot gas.
Both our observations of CHVCs and those by \citet{BBC01a} have shown
that (i) $N_{HI}$ decreases smoothly and appreciably with distance $r$
from the cloud center, and
(ii) linewidths (and hence thermal plus turbulent energy) do not 
increase with $r$.
The total gas pressure therefore decreases with $r$, as it should if the cloud 
is confined by dark matter gravity, in contrast to the constant pressure
that pressure equilibrium would produce.
The thermal conductivity of an external very hot gas is still controversial
(see \citet{NM01}), but confinement by hot gas might also heat and perturb
the outer layers of a CHVC too much.

The importance of ionized hydrogen would be less
if the \ion{H}{1} in HVCs were highly clumpy, so that much of the
overall cloud area had very low column density (see \citet{CCR00}).
For W486 and W491 we have not found any strong irregularities on the scale 
of $3\farcm2$, which is quite a bit smaller than the cloud diameters.
\citet{BB00} and \citet{BBC01a} also do not find very pronounced holes.
Thus, strong clumpiness is not likely to be important for column density
distributions in the HVCs that have been mapped to date, although 
{\em volume} density fluctuations on smaller scales cannot be excluded.

For the main portions of the extended HVC complexes, $N_{HI}$ is larger
than for CHVCs and the contribution from ionized hydrogen should be
unimportant.
On the other hand, the sources described by \citet{MLS95} and 
\citet{LMPU02}, while mostly outlyers of extended HVC complexes,
mostly have $N_{HI} << 2 \times 10^{19}\;{\rm atoms}\;{\rm cm}^{-2}$,
and ionized hydrogen is likely to dominate over the neutral phase.
A small number, up to 7\%, of these sources seem to be smaller than 
the $21\arcmin$ beam and might be small enough for the actual 
central $N_{HI}$ to be of order 
$2 \times 10^{19}\;{\rm atoms}\;{\rm cm}^{-2}$ or only slightly
smaller, similar to the three mini-HVCs described above.
Such compact sources contribute little to the total gas mass,
and even less to obscuration of UV from quasars, but their $N_{HI}$ 
distribution function is important whether the sources are isolated
or extensions of HVC complexes.
Better statistics in this range of $N_{HI}$ could pin down a global
value of the edge column density $N_{1/2}$ or find evidence for
variations in $N_{1/2}$ (see the discussion in \citet{CB02}).
The value of $N_{1/2}$ may well vary from place to place because the
flux of ionizing radiation may vary for two reasons:
(i)  More than 37\% of directions on the sky have $N_{HI}$ above 
$1.6\times 10^{17}\;{\rm atoms}\;{\rm cm}^{-2}$ (unit UV optical
depth at the Lyman edge --- \citet{MLS95}), so the extragalactic 
flux may be highly obscured in some places and less so in others;
and (ii) in some galaxy group environments the extragalactic ionizing
UV flux may be greatly enhanced by some UV leaking out of galaxies
\citep{BHFQ97, CDT02, CBF02}.

Quite apart from the CHVC distance controversy, it is puzzling why
there are so few starless clouds compared with dwarf galaxies in 
other galaxy groups if 
$N_{1/2} \sim 2 \times 10^{19}\;{\rm atoms}\;{\rm cm}^{-2}$ is 
universal:
$N_{1/2}$ is then much less than the minimum $N_{H, {\rm tot}}$ for star
formation to occur, and one should expect there to be hydrogen clouds with
$N_{H, {\rm tot}}$ a few times larger than 
$2 \times 10^{19}\;{\rm atoms}\;{\rm cm}^{-2}$, mostly neutral and easily
detected, but starless.
The puzzle could be relieved if $N_{1/2}$ were appreciably larger in many 
galaxy groups, since most clouds would
be either highly ionized or rich in stars.

\section{Conclusions and summary}

High sensitivity Arecibo mapping along a radial spoke in each of two Compact
High Velocity Clouds (CHVCs, defined as a slightly broader class than the very
compact and very isolated examplars catalogued by \citet{BB99})  
along with preliminary work on two more CHVCs
has led to the following conclusions:

\noindent
(1)  The neutral hydrogen column density falls exponentially, with 
scalelengths of $5\farcm 5$ and $7\arcmin\:$ for W486 and W491, 
respectively.
The exponential decline continues as far as we can trace the gas, 
from column densities near $2 \times 10^{19}$ to below 
$10^{18}\;{\rm atoms}\;{\rm cm}^{-2}$.

\noindent
(2)  The most reasonable explanation of these results is that the 
exponential decline we observe is the truncation edge itself, with
the edge column density $N_{1/2}$, defined as the neutral gas density where
the neutral and ionized fractions are equal, around 
$2 \times 10^{19}\;{\rm atoms}\;{\rm cm}^{-2}$, similar to $N_{1/2}$ values
for nearby galaxies and for Lyman Limit Systems.
The cores of the CHVCs we mapped are only slightly above this value.
The bulk of the gas in these CHVCs must be ionized 
rather than neutral, the scalelengths must be larger for total
hydrogen (ionized plus neutral) than for \ion{H}{1} alone, and the
baryon fraction must be larger than the \ion{H}{1} mass fraction alone.

\noindent
(3)  The cloud edges are relatively smooth on the scale of the 
$3\farcm2$ beam.
There is little sign of clumpiness and the linewidths are fairly constant.

\noindent
(4)  W491 has a superimposed mini-HVC, displaced in velocity 
by $66\;{\rm km}\;{\rm s}^{-1}$, with a peak column density only 
$5 \times 10^{18}\;{\rm cm}^{-2}$.
Preliminary data toward future mapping of two more CHVCs
has revealed two more mini-HVCs of similar size and similarly 
low central column density, displaced even more in velocity.

To date there has been very little mapping of the 
HVC population with the sensitivity and resolution required to detect 
cloudlets like the mini-HVC superimposed on W491.
Nor do we know whether the smoothness of the edges of W486 and W491 is typical 
of the general CHVC population.
Mapping of a statistically significant sample of CHVCs, 
with further exploration around the edges of each, will be 
required before we know whether or not there are differences in the 
corrugation of the edges of more-isolated and less-isolated CHVCs that might 
indicate different exposures to Galactic and extragalactic UV.
Mapping of some sightlines independent of known CHVCs
with sufficient sensitivity to detect low column density gas is also needed
to assess whether the three mini-HVCs really are independent of the
CHVCs against which they are seen.
The few isolated beam positions at which low column density emission
was found by \citet{LMPU02} might be fruitful starting points.
It will also be important to map some sightlines further removed 
from the Magellanic Stream to ascertain that the mini-HVCs are not 
related solely to that tidal feature.
Arecibo resolution is important since the mini-HVCs we have detected to date
subtend only a few Arecibo beams (and only a single Green Bank Telescope
beam).
A statistical sample of mini-HVCs will be needed before we can assess
(a) the likelihood that they are candidates for the Local Group wide infalling 
gas clouds for which \citet{BSTHB99} (and others before them) have argued,
and (b) implications for the total hydrogen column density distribution
\citep{CB02}.

\acknowledgments
We thank R. Braun, J. Charlton, E. Corbelli, M. Shull and the anonymous referee 
for valuable discussions and comments, and we are grateful to J. Dickey and 
S. Stanimirovi\'{c} for making ANZUV available to us.
P. Perilat provided invaluable assistance with procedures at the 
telescope, and the Arecibo telescope operators were, as always, 
courteous and capable.
R. Gildea and A. Hirani assisted with the Arecibo observations.
Partial funding was provided by the Crafoord Prize Fund at Cornell University.
Travel support was provided by NAIC and a grant from Lafayette College.

\begin{figure}
\plotone{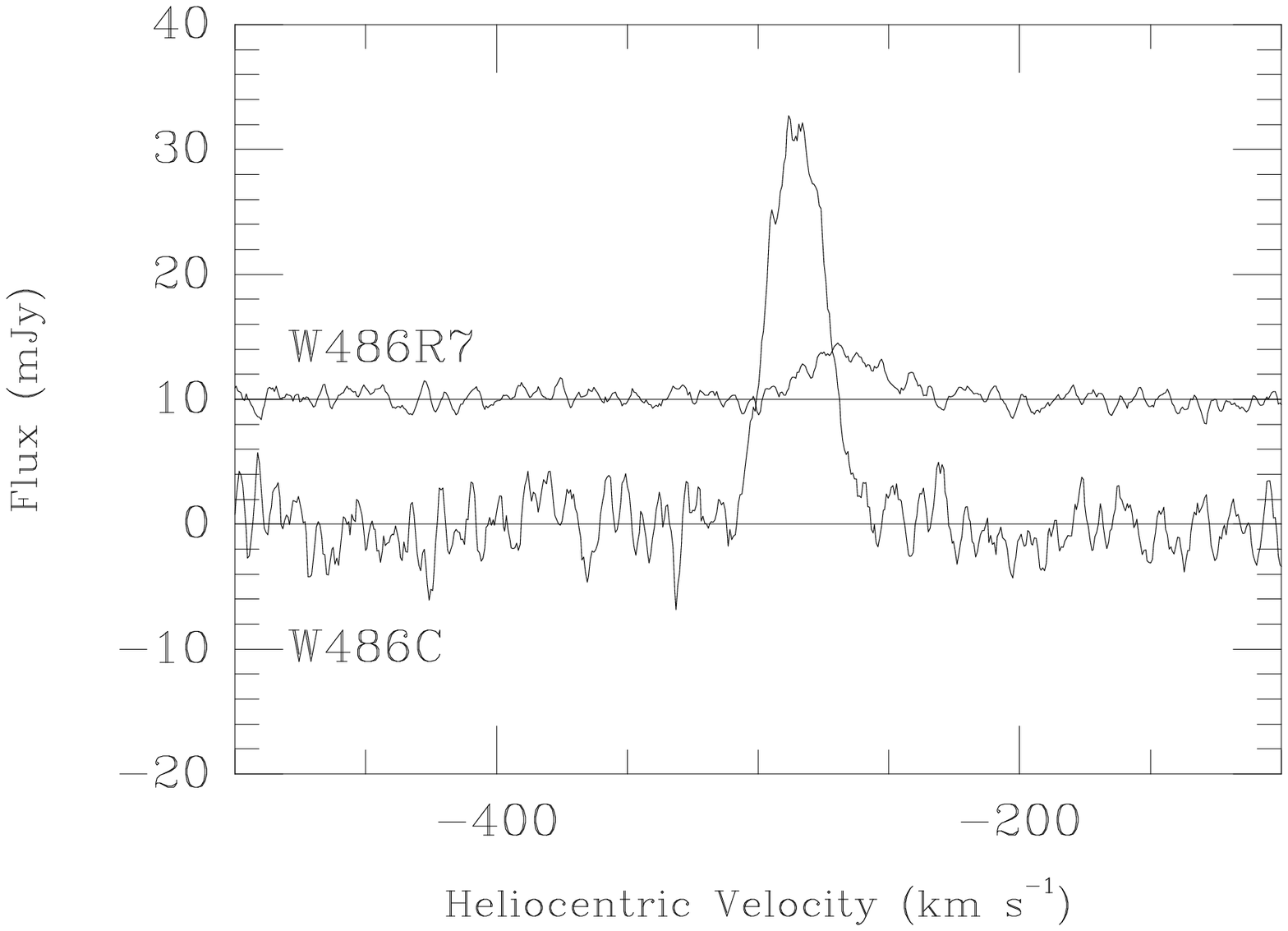}
\caption{
Arecibo spectra for the core of W486 (labelled W486C) and for a point 
(labelled W486R7)
near the end of the mapped spoke, $7 \times 3.\arcmin2$ out from the core.
The latter has been displaced vertically by 10 mJy for clarity.}
\end{figure}

\begin{figure}
\plottwo{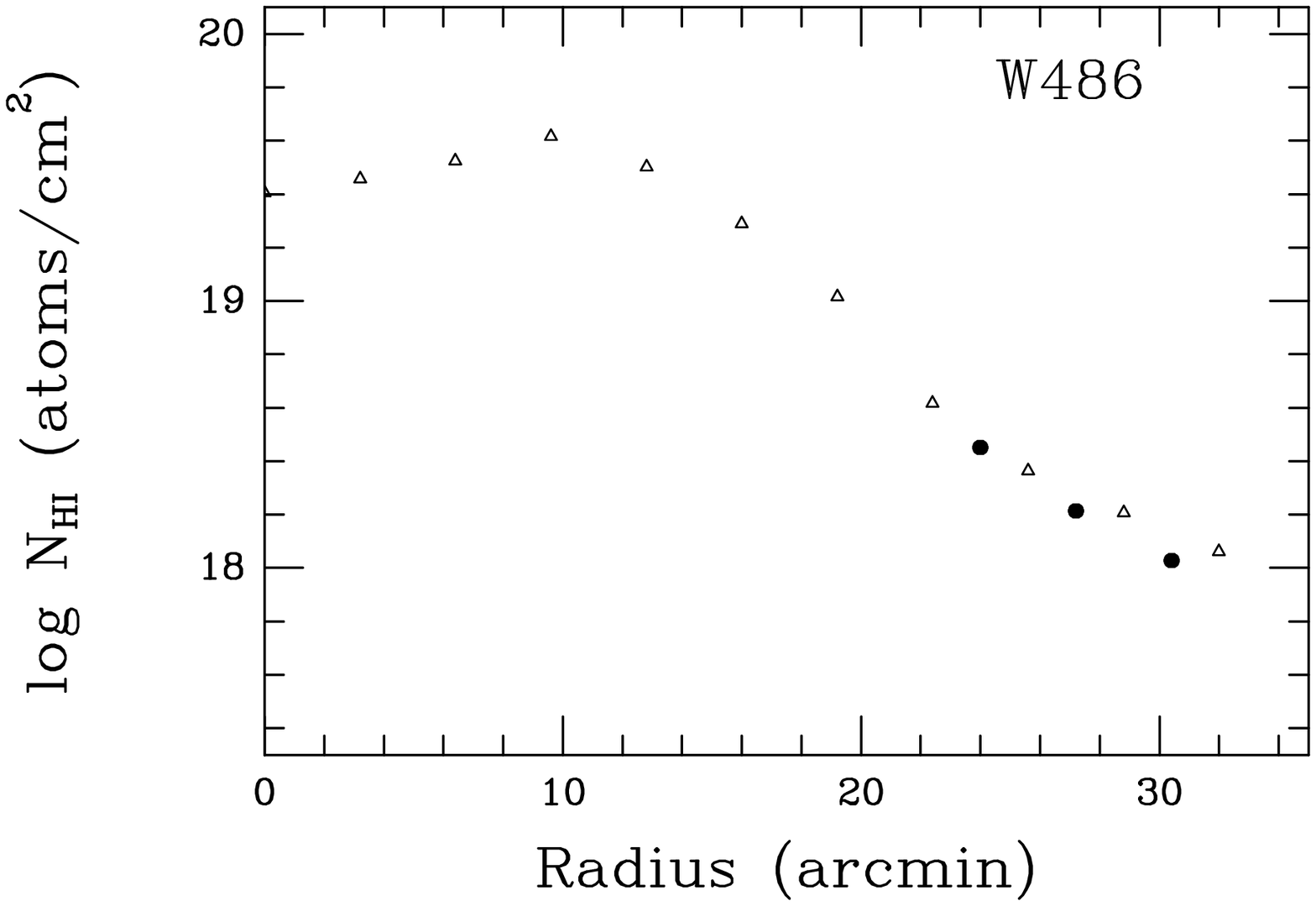}{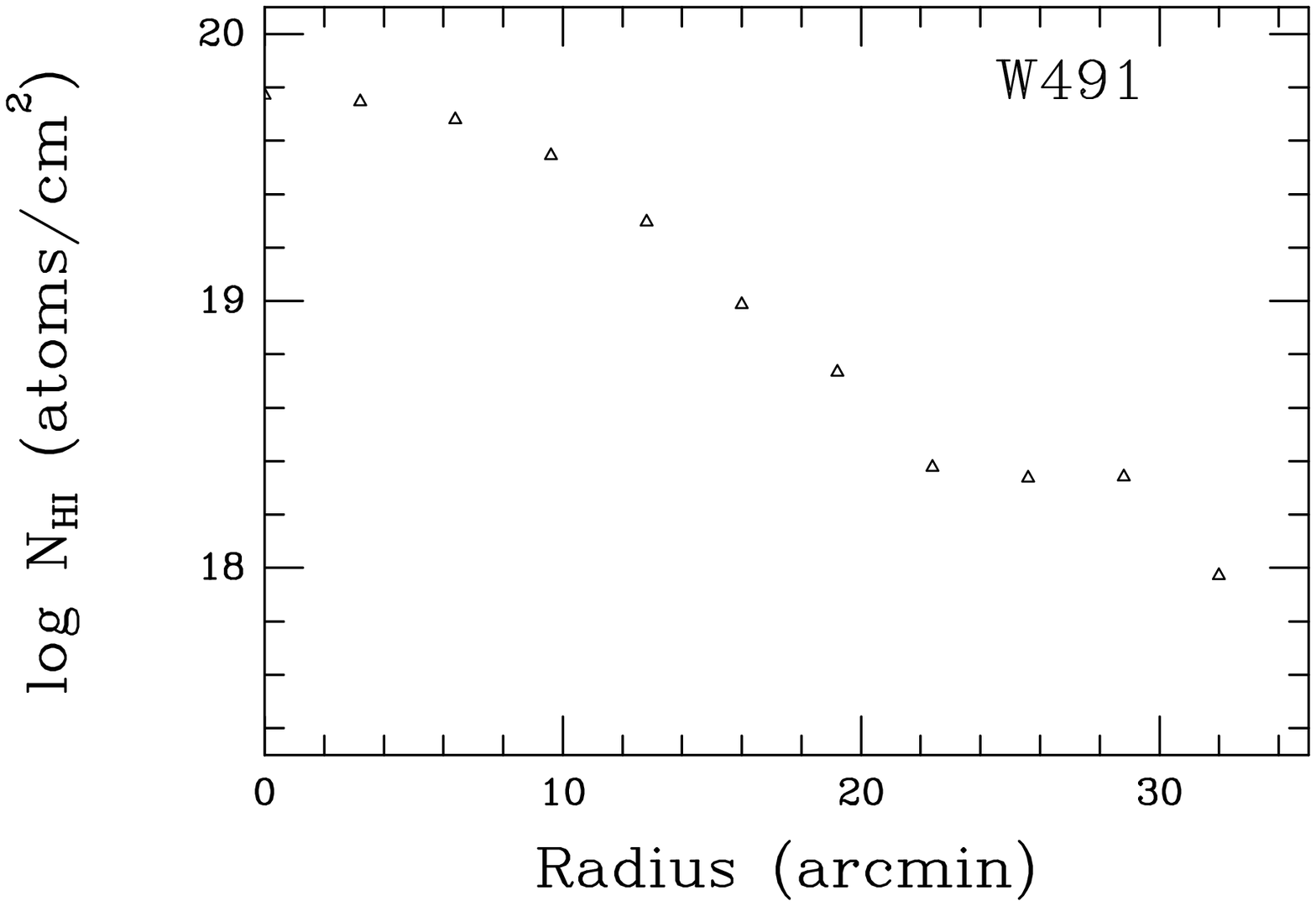}
\caption{
Neutral hydrogen column density vs.\ radius for the main spokes 
of W486 (left panel) and W491 (right panel).  
For W486, the open triangles are single beam results.  
Solid circles are four-beam averages (two adjacent beams along the spoke, 
and one beam on either side).
No corrections for sidelobes have been made, but rough estimates indicate
that each point would shift downward by 0.10 in $\log N_{HI}$ without
significantly affecting the measured scalelengths.}
\end{figure}

\begin{figure}
\plotone{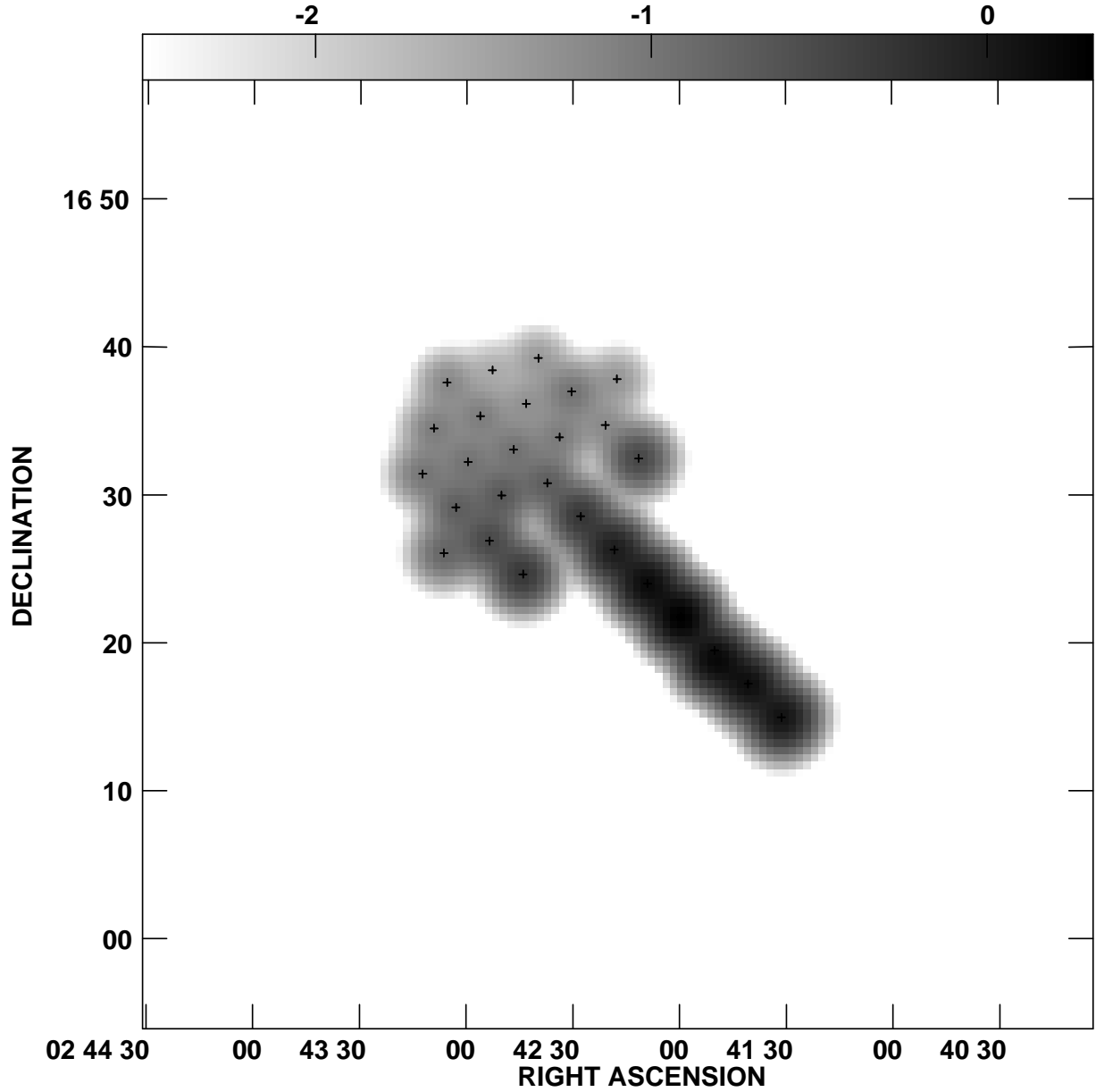}
\caption{
Logarithmic greyscale representation of the Arecibo integrated fluxes for W486.
Each observed point has been convolved with a 3.2 arcmin beam, and the 
centers of the observed beams are indicated by crosses.
The wedge indicates the logarithm of the ratio of column density to maximum
column density.
The greyscale range is equivalent, approximately, to a column density 
range from $6 \times 10^{17}$ to $4 \times 10^{19}$~atoms~${\rm cm}^{-2}$.}
\end{figure}

\begin{figure}
\plottwo{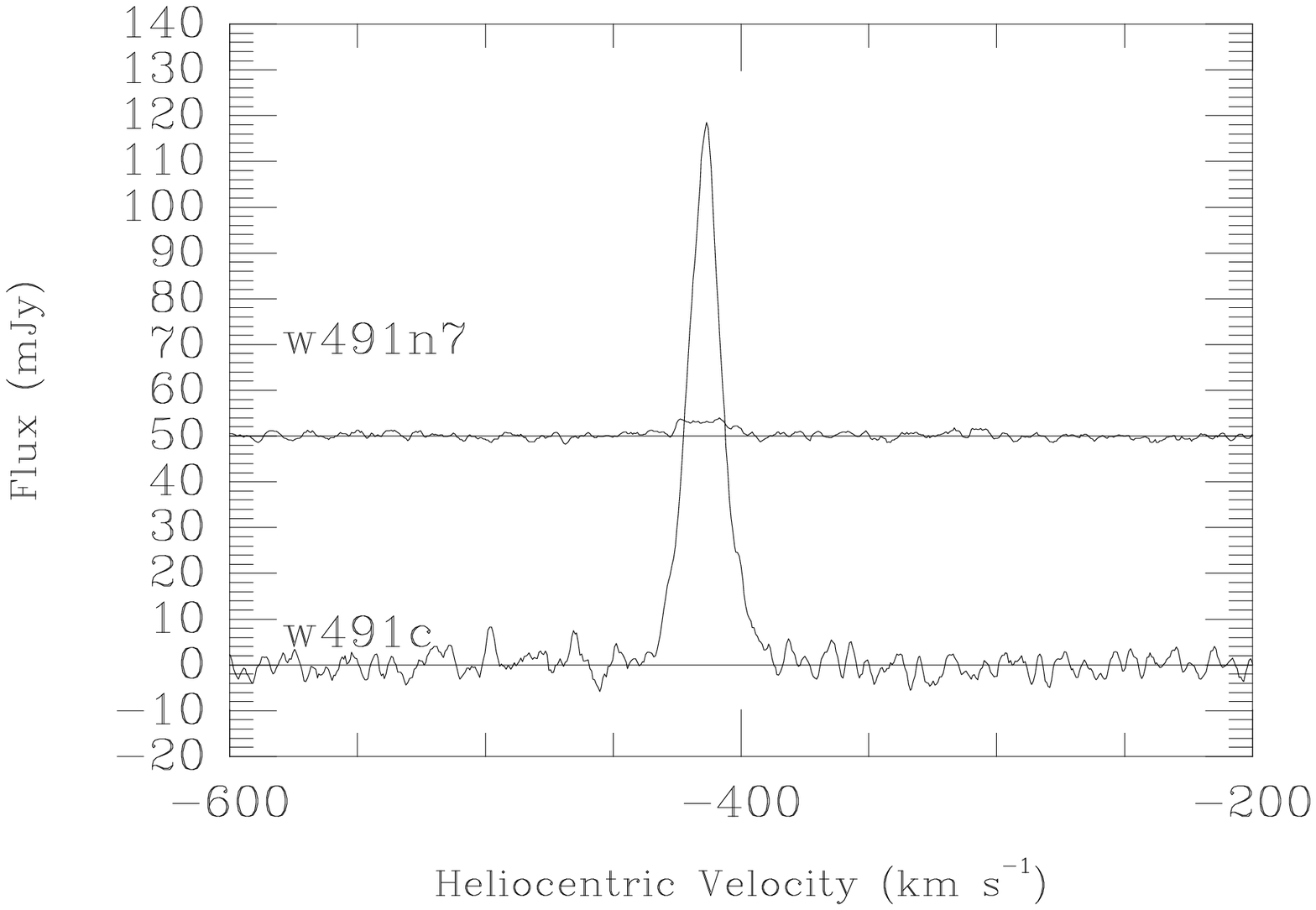}{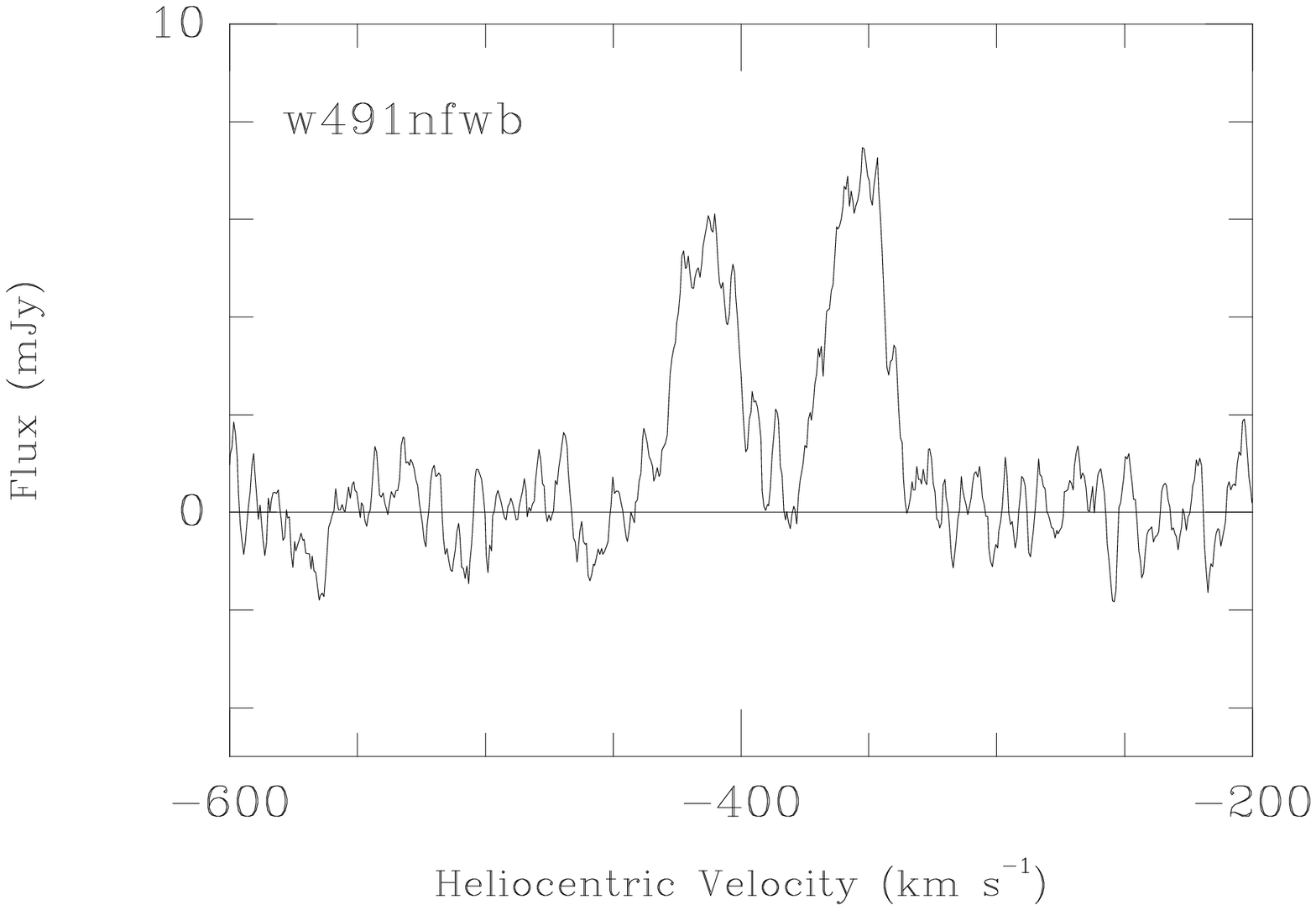}
\caption{
Arecibo spectra (lefthand panel) comparing the core of W491 
(labelled w491c) and for a point near the end of the original 
mapped spoke, $7 \times 3.\arcmin2$ out from the core.
The latter has been displaced vertically by 50 mJy for clarity.
The righthand panel (labelled w491nfwb), shows a spectrum from a beam 
near the center of the mini-HVC, which is the feature at the higher
velocity.}
\end{figure}

\begin{figure}
\plottwo{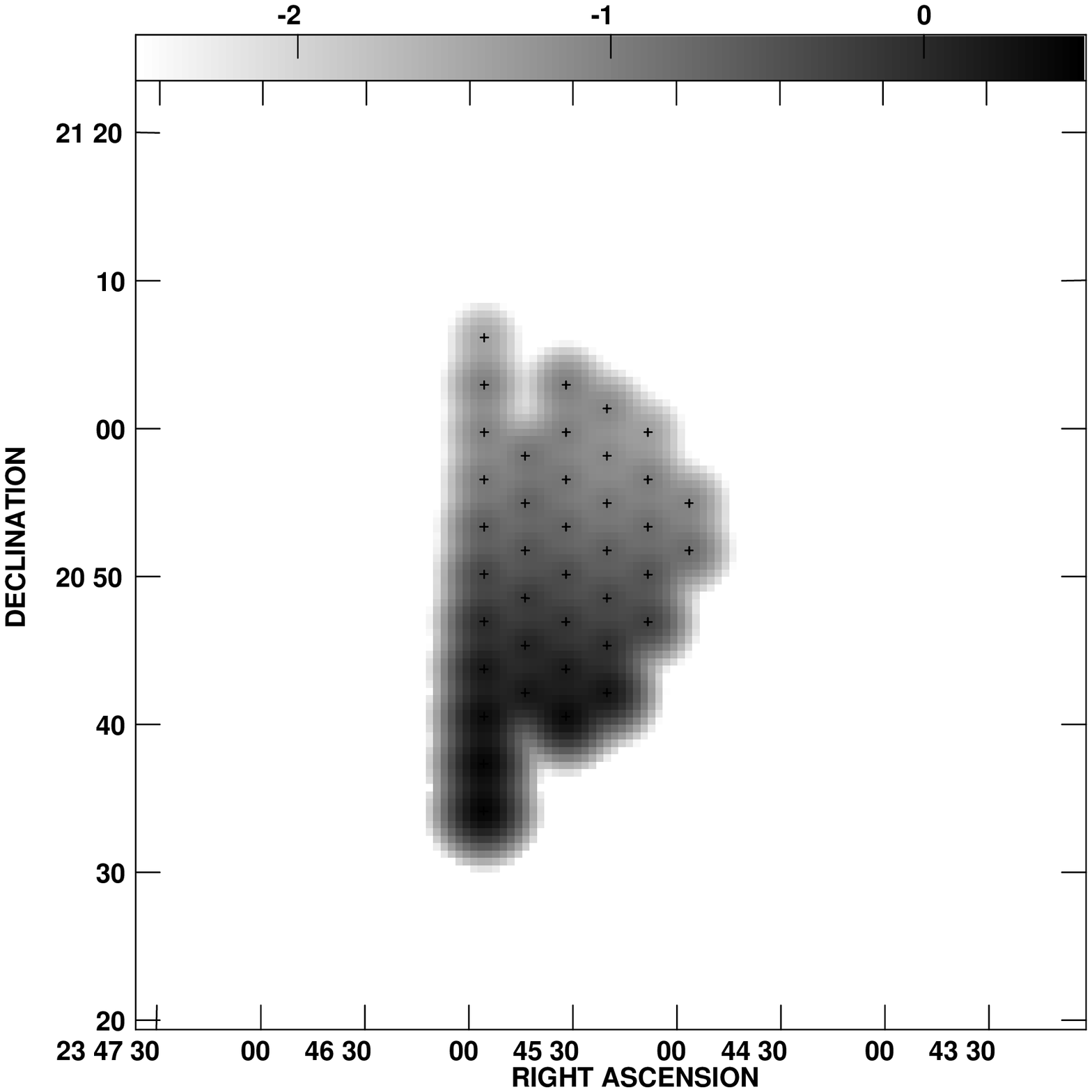}{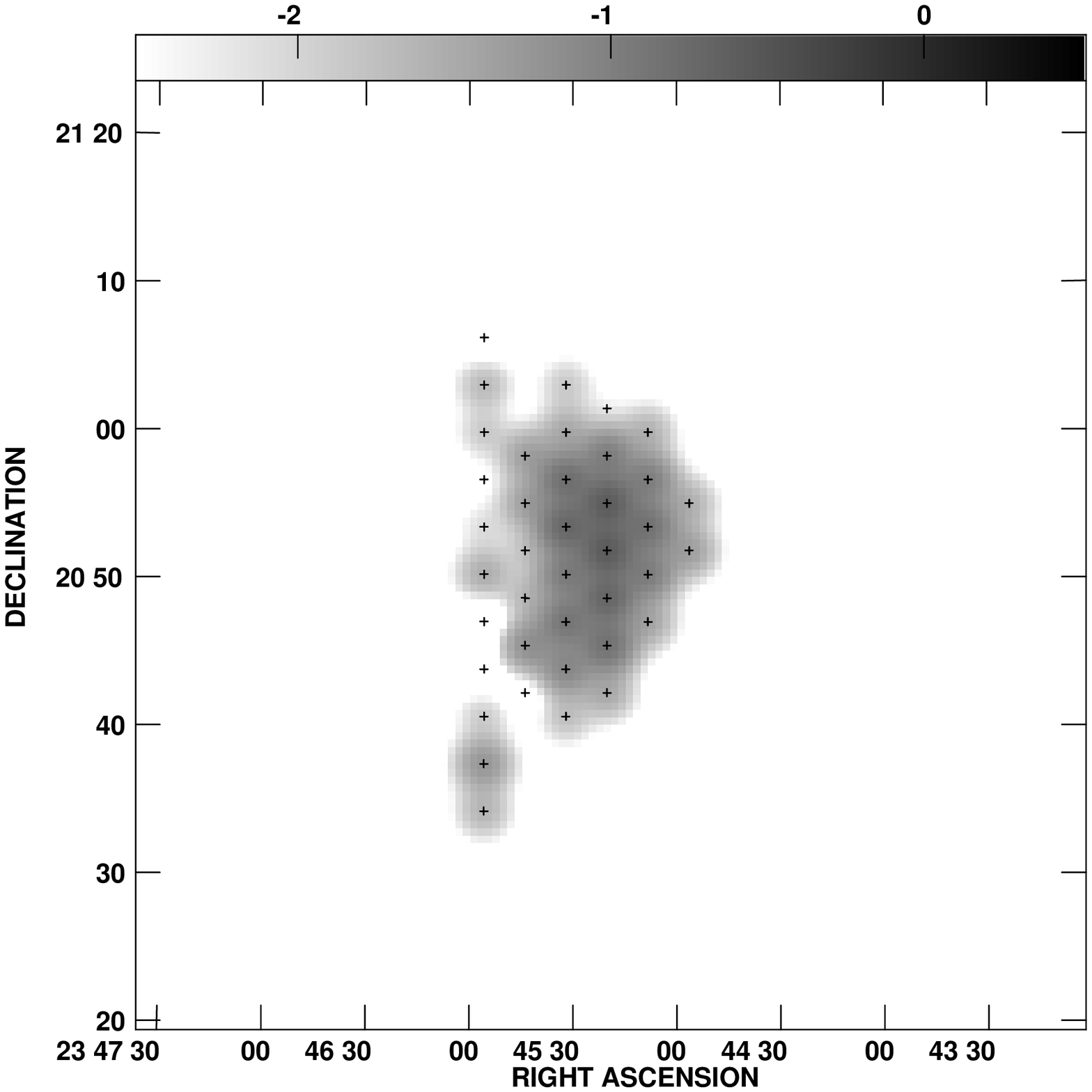}
\caption{
Logarithmic greyscale representation of the Arecibo fluxes for W491, 
integrated over the velocity 
range $-440$ to $-380\;{\rm km}\:{\rm s}^{-1}$ appropriate to the main cloud 
in the left panel, and over the range $-380$ to $-320\;{\rm km}\:{\rm s}^{-1}$ 
appropriate to the mini-HVC in the right panel.
Each observed point has been convolved with a 3.2 arcmin beam and the 
beam centers are indicated by crosses.
The wedge indicates the logarithm of the ratio of column density to maximum
column density.
The greyscale range is equivalent, approximately, to a column density 
range from $6 \times 10^{17}$ to $6 \times 10^{19}$~atoms~${\rm cm}^{-2}$.}
\end{figure}

\begin{figure}
\plotone{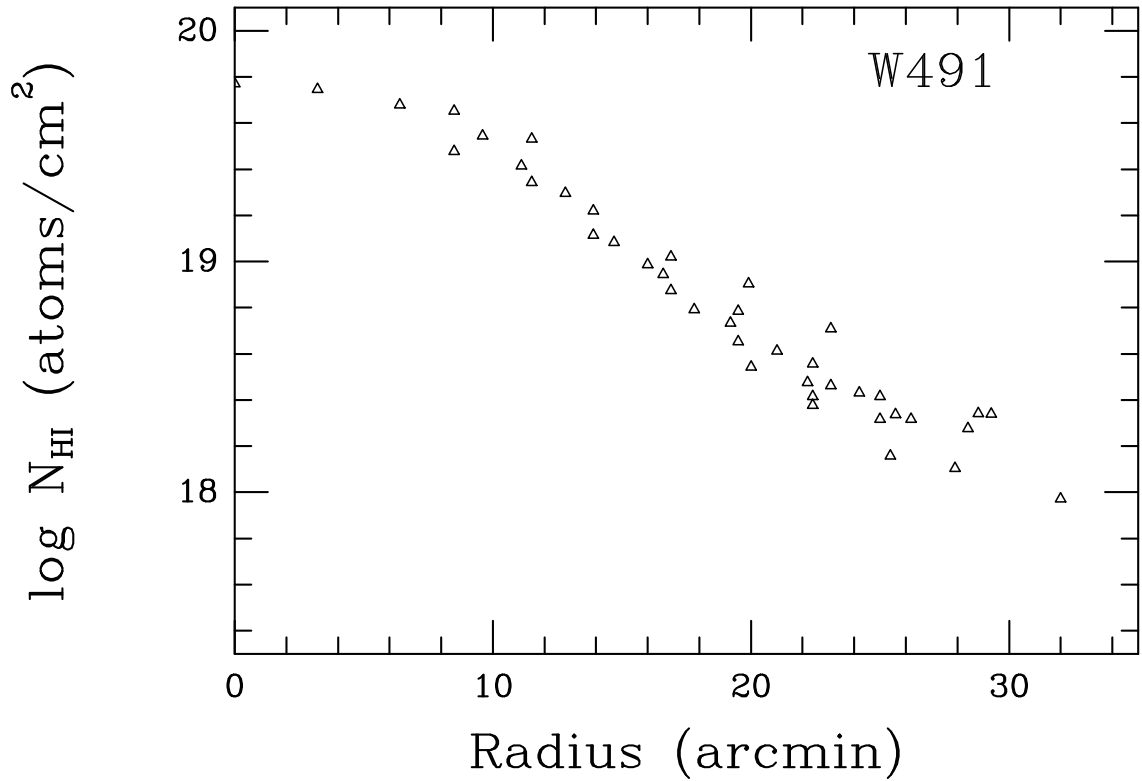}
\caption{
Neutral hydrogen column density vs.\ radius for all 
observed points on the main cloud 
(integrated over the velocity range $-440$ to 
$-380\;{\rm km}\:{\rm s}^{-1}$) of W491, plotted as a function of the 
distance from the nominal center.
}
\end{figure}

\begin{figure}
\plotone{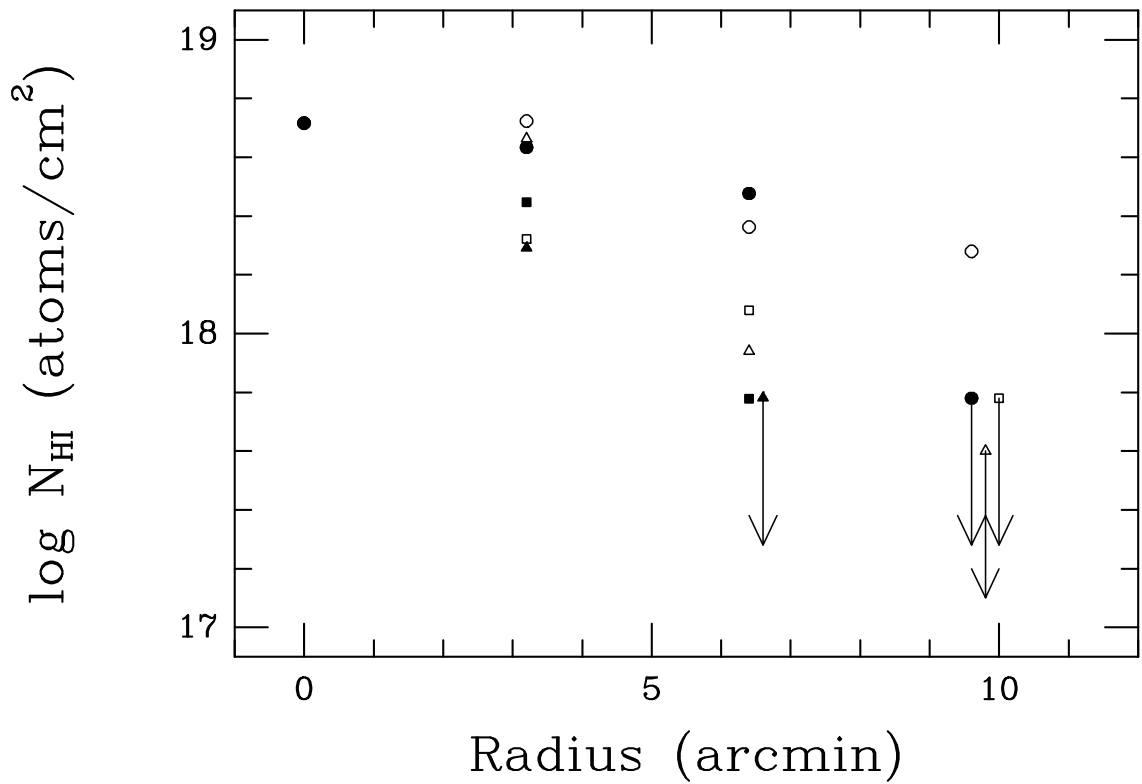}
\caption{
Neutral hydrogen column density vs.\ radius for the mini-HVC 
superimposed on W491.
Different symbols represent spokes at different position angles 
(all measured E from N):
$0\arcdeg\:$ (open circles), $60\arcdeg\:$ (open triangles), 
$120\arcdeg\:$ (open squares), $180\arcdeg\:$ (solid circles), 
$240\arcdeg\:$ (solid triangles), and  $300\arcdeg\:$ (solid squares).
Upper limits are denoted by the downward arrows.
Sidelobe corrections have not been made, but should be minimal.}
\end{figure}

\begin{figure}
\plotone{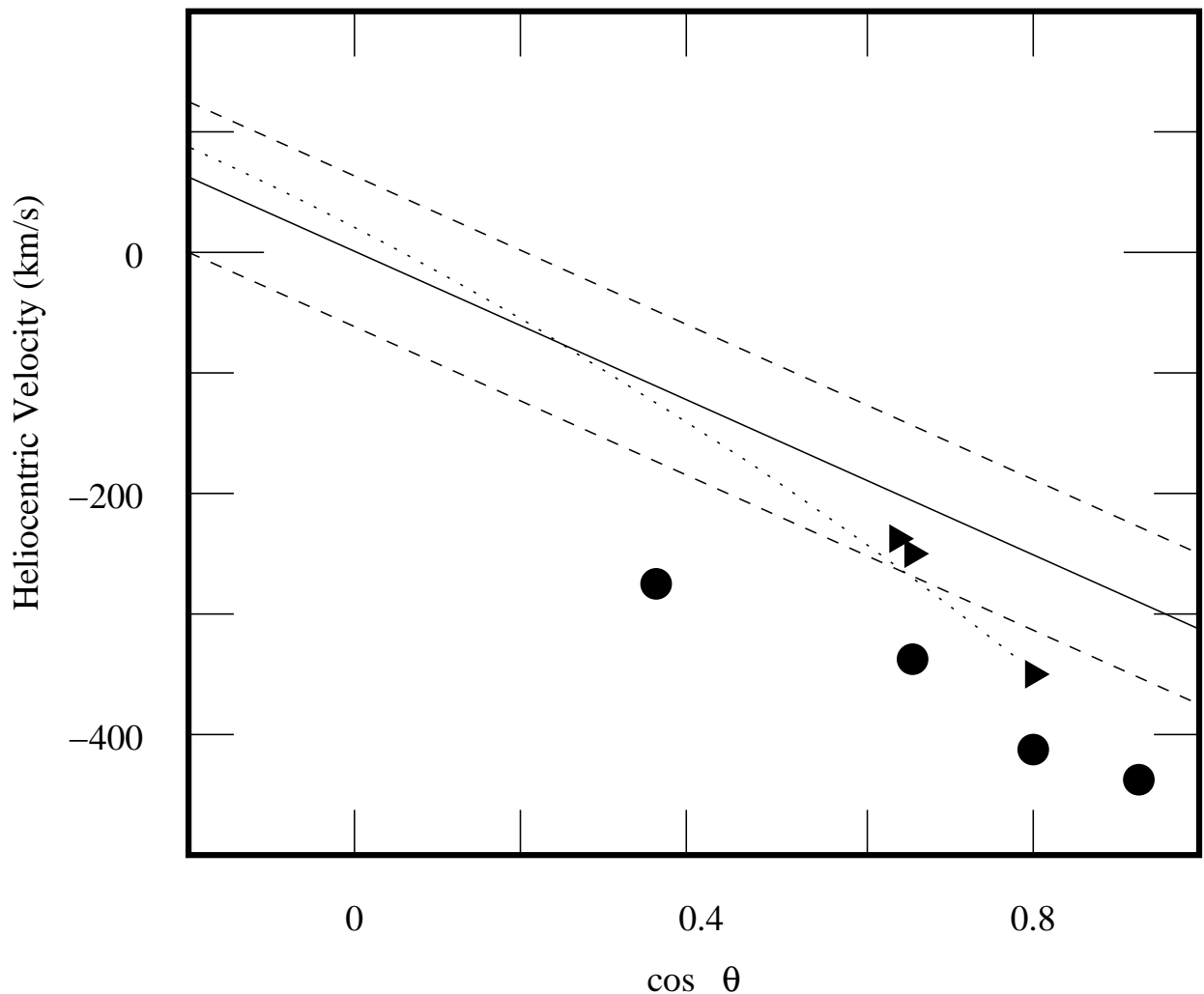}
\caption{
Heliocentric velocities of CHVCs plotted against the cosine of the 
angle $\theta$ from the barycenter of the Local Group.
The four HVCs we have mapped at Arecibo (W486 and W491 reported here,
W413 and W479 to follow in the future) are shown as solid circles,
and the three mini-HVCs as solid triangles.
The solid and dashed lines show the solar motion toward the barycenter 
of the Local Group with one standard deviation of velocities of 
Local Group member galaxies above and below, as in Fig.\ 5 of \citet{BB99}.
The Magellanic Stream is shown (approximately) as a dotted curve.}
\end{figure}

\end{document}